\documentclass[11pt,a4]{article}
\usepackage{moriond,epsfig,graphicx}
\usepackage{graphics}
\def\be{\begin{equation}} 
\def\ee{\end{equation}} 
\def\bea{\begin{eqnarray}} 
\def\eea{\end{eqnarray}} 
 
\def\beq{\begin{equation}}
\def\eeq{\end{equation}}
\def\bea{\begin{eqnarray}}
\def\eea{\end{eqnarray}}
\def\bq{\begin{quote}}
\def\eq{\end{quote}}

\def\gappeq{\mathrel{\rlap {\raise.5ex\hbox{$>$}}
{\lower.5ex\hbox{$\sim$}}}}
\def\lappeq{\mathrel{\rlap{\raise.5ex\hbox{$<$}}
{\lower.5ex\hbox{$\sim$}}}}

\def\bea{\begin{eqnarray}}   
\def\eea{\end{eqnarray}}

\begin{document}\vspace*{2cm}
\begin{flushright}
\texttt{CI-UAN/04-02fT}\\
\texttt{LPT-Orsay/04-31}\\
\texttt{hep-ph/XXXXXXXX} 
\end{flushright}
\vskip 5pt
\title{Leptogenesis at the TeV scale} 
\author{\sf Asmaa Abada $^{1}$, \underline{\sf Habib Aissaoui}
\footnote{Presented the invited talk.} $^{1,2}$, \sf Marta Losada $^{3}$}
\address{$^1$ Laboratoire de Physique Th\'eorique, Universit\'e de Paris XI,\\ 
B\^atiment 210, 91405 Orsay Cedex, France \\ $^2$ Laboratoire de Physique 
Math\'ematique et Physique Subatomique,
\\ Universit\'e Mentouri, Constantine, 25000, Algeria.\\ 
 $^3$ Centro de Investigaciones, Universidad Antonio Nari\~{n}o,
 \\Cll. 58A No. 37-94, Santa Fe de Bogot\'{a}, Colombia}
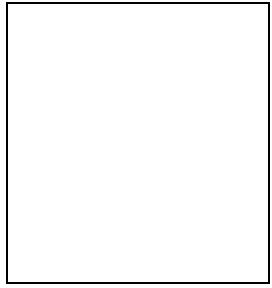
\maketitle
\abstracts{We consider a generic model with four gauge singlets which generates 
successfully the right amount of baryon asymmetry through leptogenesis at the
 TeV scale. It also reproduces  left-handed neutrino masses compatible with
  present data.\\In the present work, we perform a study of thermal production
   mechanisms of the lightest right-handed neutrino $N_{1}$ responsible for the lepton
    asymmetry taking into account all dominant processes (decays,  $\Delta L = 1$ and 
    $\Delta L=2$ scatterings) by solving the full Boltzmann equations.} 

\section*{Introduction}
 The baryon asymmetry of the universe (BAU) plays a special role in both 
particle physics and cosmology. 
This fact is usually expressed in terms of the ratio of the baryon number 
density $n_B$ to the entropy $s$ of the universe. 
The latest measure from WMAP for this ratio is 
determined to be \cite{WMAP}
\bea
\eta=\frac{n_B-n_{\bar B}}{s}\simeq\frac{n_B}{s}={{6.5}^{+0.4}_{-0.3}} 
\times 10^{-10} \ .\eea

A viable model of baryogenesis must fulfil Sakharov's conditions \cite{sakharov},
 namely it must contain: \begin {itemize}\item {Baryon number violation}\item {$C$ 
 and $CP$ violation}\item {A deviation from thermal equilibrium.} \end{itemize} 
 It is well known that in the Standard model (SM) the baryonic ($B$) and leptonic 
  ($L$) numbers are violated at the quantum level and that there is $CP$ violation 
 in the quark sector.  However, it was established that the asymmetry 
 created in this sector
is not enough to explain the  observed BAU, see for example 
 \cite{smcp}.\\ It is now established that neutrinos  have non vanishing masses 
  (Superkamiokande \cite{superk}, SNO \cite{sno}, ...).    One of the most interesting 
  consequences of this, is the possibility to generate the BAU 
 through leptogenesis 
  via the non
equilibrium decay  of heavy right-handed (RH) neutrinos \cite{FY}. 
  Including such heavy particles gives us the possibility to generate natural 
  small masses to the neutrinos via the seesaw mechanism \cite{seesaw}, 
  and leads to $CP$ violation in the leptonic
sector \cite{roulet}. 
  This leptonic asymmetry is partially converted into a baryonic asymmetry through 
  non perturbative effective 
  sphaleron interactions \cite{kuzmin}. It was shown that a lower bound on the mass 
  of the added RH neutrinos can be 
  placed $M\sim{10^{8}}$ GeV \cite{bdp1,di} for a minimal extension of the SM with 
  3 RH neutrinos. \\
  Nevertheless, it is expected that new physics will appear not so far from the TeV 
  scale. So, considering a model of leptogenesis at this scale can be an interesting 
  alternative, and can be indirectly tested in upcoming experiments. 
  The author of ref. \cite{hambye} has illustrated some difficulties which can plague
   leptogenesis models at the TeV scale, namely:  a) obtaining adequate values of the 
   lepton asymmetry given the constraint from
 the out-of-equilibrium decay,  b) damping effects if the decaying particle has gauge 
 interactions,  c)an extremely small value of the neutrino masses, or 
  d)it is difficult to have a model in which the same interactions produce the
   asymmetry and non-zero neutrino masses. The author also discusses possible
    enhancement mechanisms which could overcome the above mentioned difficulties
    . Other possible TeV scale models  have been also discussed in 
    Refs. \cite{Senami,Senami2,Boubekeur}. \\
    In the present work, by solving Boltzmann equations (BE) for  a generic
     model proposed in ref.\cite{asm}, we perform a study of the  thermal
      production mechanism related to the lightest RH neutrino inducing 
      the leptonic asymmetry. In section 2, we present the model and its m
      ain features, in section 3, we solve the BE for, both a toy model with 
      only two generations, and our generic model with four gauge singlets,
       and finally in section 4 we present our conclusions. 


\section*{The Model}
The model proposed in  \cite{asm} is based on the SM  with a minimal extension
 by adding a fourth generation which satisfies the LEP constraints and four gauge singlets $\nu_R$.  
 The Lagrangian of the model is given by \cite{asm}  
\beq L = L_{SM} + \bar{\psi}_{R_{I}} i \partial \! \! \! /~\psi_{R_{I}} -
\frac{M_{N_{I}}}{2}(\bar{\psi}_{R_{I}}^{c} \psi_{R_{I}} + h.c.) - 
(Y^{\nu}_{IJ} \bar{L}_{J} \psi_{R_{I}} \phi + h.c.) \ ,\label{lag} \eeq where $
\psi_{R_{I}}$ are two-component spinors describing the RH neutrinos 
and we define a  Majorana 4-component spinor, $N_{I} =  \psi_{R_{I}} + 
\psi_{R_{I}}^{c}$. Our index I runs from 1 to 4. The fourth component of $L_I$
corresponds to a left-handed (LH) lepton doublet which must satisfy the LEP
constraints from the Z-width on a fourth LH neutrino~\cite{pdg}. The 
$Y_{IJ}$ are Yukawa couplings and  the field $\phi$ is the (SM) Higgs boson
doublet whose vacuum expectation value is denoted by $v$.
We work in the basis in which the mass matrix for the RH neutrinos
$M$  is diagonal and real,
\beq M = diag(M_1,M_2,M_3,M_4) \eeq and define $m_D = Y_{\nu} v$. The
neutrino mass matrix for the LH neutrinos is given by,
 \beq m_{\nu}  = m_D^T M^{-1} m_D = Y_{\nu}^{\dagger} M^{-1} Y_{\nu} v^2\ .\eeq 
It is clear here that the LH neutrinos get small masses via the seesaw mechanism when 
the RH neutrinos masses are large or alternatively  for $M_i \sim$ TeV
 for values of the neutrino Yukawa couplings on the order of the electron
  Yukawa coupling.\\ The $CP$ asymmetry arises from the interference
   between tree level  and one-loop diagrams (self-energy and vertex corrections),
    and the $CP$ violation parameter is given by (for more details see ref.\cite{roulet})
    
\beq \epsilon_I = \frac{1}{(8\pi)}\frac{1}{  [Y_{\nu}^{\dagger} Y_{\nu}]_{II}}
\sum_J {\mathrm Im}{ [Y_{\nu}^{\dagger}Y_{\nu}]^2_{IJ}}
\left[f\left(\frac{M_{J}^2}{M_{I}^{2}}\right) + g\left(\frac{M_{J}^2}{M_{I}^{2}}\right)
\right] \label{asym}\ , \eeq 
where \bea f(x) &=& \sqrt x[1 - (1+x)\ln\frac{1+x}{x}],
 \nonumber\\ g(x) &=&\frac{\sqrt x}{1-x}. \eea 
 The ratio $\eta=Y_B=\frac{n_B}{s}$ is related to leptonic asymmetry ratio 
 $Y_L=\frac{n_L}{s}$ by \beq	Y_B=- \left(\frac{8N_{F} + 4N_{H}}{22N_F +13N_H}\right) 
 Y_{L}, \eeq where $N_H$ is the number of Higgs doublets and $N_F$ is the number 
  of fermionic families.  


\section*{Boltzmann Equations}
 The production of a baryonic or leptonic asymmetry is an out-of-equilibrium  process
  which is usually analysed using Boltzmann Equations \cite{kolb}. 
  The main processes in the thermal bath of the early universe are decays, inverse decays 
  of the RH neutrinos, and the lepton number violation $\Delta L=1$  and $\Delta L=2$,
  Higgs and RH neutrinos exchange scattering processes, 
  respectively.\\ The first BE, which corresponds to the evolution of the abundance  
   \footnote{We suppose that the asymmetry is due only to the decay of the lightest
    RH neutrino $N_1$.} of the 
lightest RH neutrino $Y_{N_1}$ involving  the decays,  
    inverse decays and $\Delta L=1$ processes is given by 
\bea{dY_{N_{1}}\over dz}=-{z\over sH(M_1)}
({Y_{N_{1}}\over {Y^{eq}_N}} -1)(\gamma_{_D}+\gamma_{_S}) \label{be1}\, \eea
where $z=M_1/T$ and  $\gamma_{_D},\gamma_{_S}$ are the interaction rates for the decay 
and scattering  $\Delta L=1$ contributions, respectively. \\ 
The BE for the lepton asymmetry is given by
 \bea {dY_L\over dz}=-{z\over sH(M_1)}
 \bigl[\epsilon_1 \gamma_{_D}( {Y_{N_{1}}\over Y^{eq}_N}-1)
 + \gamma_{_W} {Y_{L}\over Y^{eq}_L}\bigr]\label{be2}\, \eea 
 where $\epsilon_1$ is the $CP$ violation parameter given by eq.(\ref{asym})
  and $\gamma_{_W}$, which is  function of $\gamma_{_D}$ and $\gamma_{_S}$ and 
  $\Delta L=2$ interaction rate processes, is a washout factor responsible 
  for the damping of the produced asymmetry.  Explicit expressions  of the interaction 
   rates $\gamma_{_D}$ ,  $\gamma_{_S}$ and $\gamma_{_W}$
    can be found  in for example refs. \cite{plumacher,branco} 
    (and references therein) where one can easily extend them to our model. 

\subsection*{Toy Model} In ref. \cite{asm}, the case of two generations as a toy model 
was considered and it was estimated that  this toy model fits the right amount of $Y_B$. 
In this section, we solve eqns. (\ref{be1}) and (\ref{be2}), for one possible texture for the neutrino Yukawa coupling matrix. 
This texture allows us to
illustrate the interplay of the different terms which contribute to the $CP$
asymmetry in eq. (\ref{asym}), and has the form 
\beq Y_{\nu} = \left(\begin{array}{cc} \epsilon & \epsilon\\ \alpha & 1\\
\end{array}\right). \eeq 
\begin{figure}[t] 
\begin{center}
\includegraphics[width=12cm]{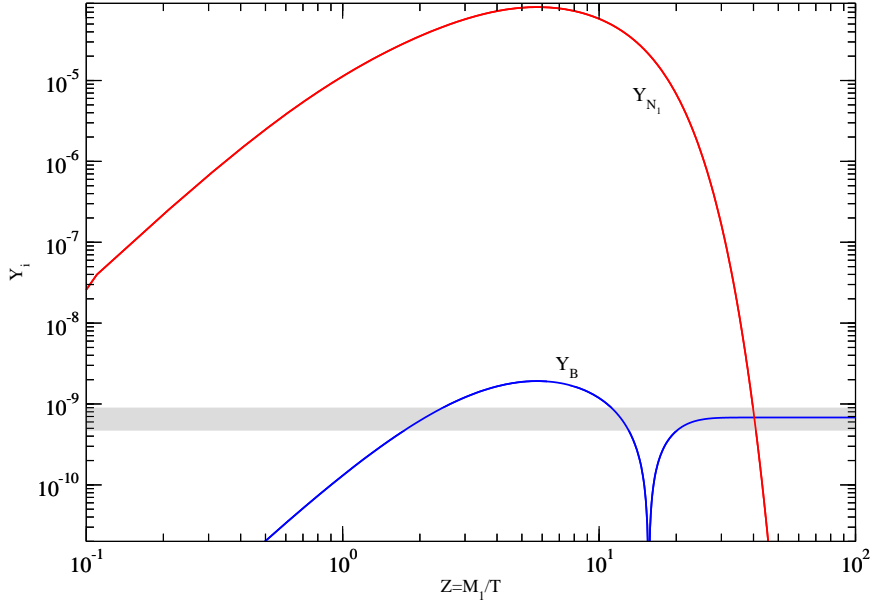}
\caption{Toy model abundance $Y_{N_1}$ and the baryon asymmetry $Y_B$.} \label{toy} 
\end{center}
\end{figure}
 In figure \ref{toy} we plot the integration of the BE for the  abundance  $Y_{N_1}$ 
 and the baryon asymmetry $|Y_B|$ as a function of $z=M_{1}/T$ for the values of 
 $M_1=450$ GeV, 
 $M_2 = 650$ GeV, $\epsilon = 1.6 (1-i)\times 10^{-11}$ and 
 $\alpha= 1.6\times 10^{-7}(1-10^{-4}i)$. We are choosing 
 these values for the sake of illustration, as they  allow us to satisfy the constraints on neutrino masses.
For instance, applying the seesaw mechanism  with the same set of parameters, 
we obtain a heavy left-handed
neutrino with a mass above $45$ GeV, as expected, and a light one with a mass of  the
order of $10^{-4}$ eV. Finally, the generated  values for the $CP$ parameter and the final baryon asymmetry are
$|\epsilon_1|\simeq 2.5\times 10^{-5}$ and $\eta_B \simeq 6.7\times 10^{-10}$, respectively.

\subsection*{Four generations}We now generalise the texture $Y_{\nu}$ of the toy model 
to the case of 4 generations~:\bea Y_{\nu} = C \left(\begin{array}{cccc} 
\epsilon & \epsilon & \epsilon &\epsilon\\ \epsilon & 1 & 1 & 0 \\ \epsilon & 1 & 1 & 0
 \\ \alpha & 0 & 0 &1/C\\ \end{array}\right). \eea

\noindent This will induce to first order the following  mass matrix for the light 
LH neutrinos

\bea m_{\nu} =  C^2 \frac{v^2}{M}\left(\begin{array}{ccc} \epsilon^2 &\epsilon & \epsilon \\ \epsilon & 1 & 1 \\ \epsilon & 1 &
1\end{array}\right),
\eea 
which is a simple form of the light neutrino mass matrix which can account for all data \cite{yanagida,Ramond2}. $C$ is a small number that makes $C^2
v^2/M$ to be of the correct order of magnitude.

\begin{figure}[t] \begin{center}
\includegraphics[width=12cm]{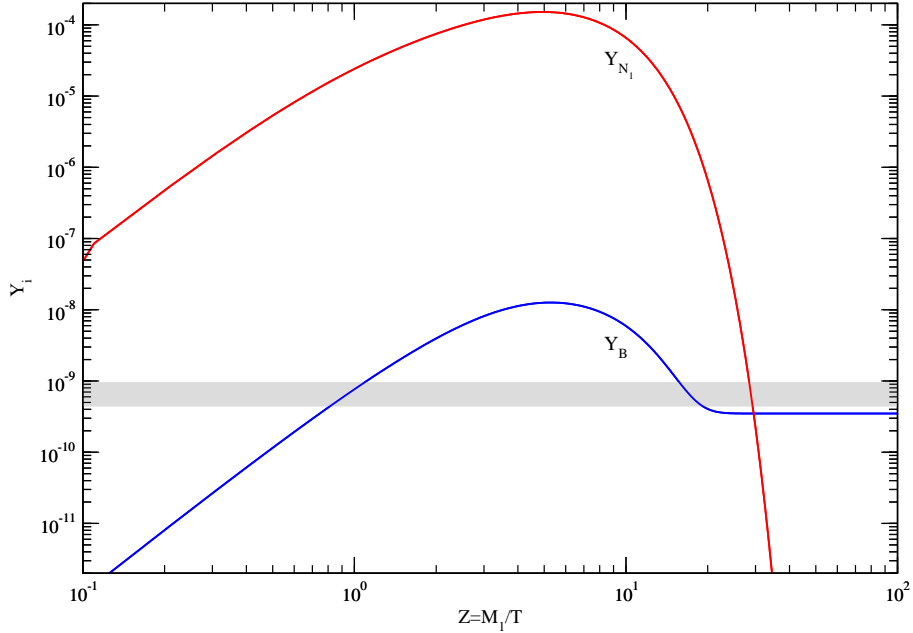} \caption{Abundance $Y_{N_1}$ and the baryon asymmetry $Y_B$ for the four generation model.} \label{model}\ 
\end{center}
\end{figure}
 In figure \ref{model} we plot the solutions of the BE as a function of $z=M_{1}/T$ for the
 following set of parameters, chosen for illustration : $M_1=450$ GeV, 
$M_2=2\times 10^5$ GeV, $M_3=10^6$ GeV $M_4 = 650$ GeV, 
$\epsilon = 2.4 \times 10^{-4}(1-.01 i)$, $\alpha= 1.2\times 10^{-3}(1-i)$ and 
$C=5\times 10^{-5}$. 
Applying the seesaw mechanism to our model for the same values of the parameters 
 we obtain a heavy left-handed neutrino with a mass above $45$ GeV, and the
  light neutrinos with masses of the 
order of $10^{-1}$ eV to $10^{-7}$ eV.
   The generated values for the $CP$ and the baryon asymmetries  obtained for 
   this texture are $|\epsilon_1|\simeq 6.8\times 10^{-5}$ and $\eta_B \simeq
    3.4\times 10^{-10}$ respectively.

\section*{Conclusion}
In this work, we have analysed the possibility of leptogenesis at the TeV scale
 for a generic model. By solving the BE equations for the production of the heavy RH
  Majorana neutrinos and the lepton asymmetry  at this scale, we  have shown that
   lepton (baryon) asymmetry can be produced. This model also  satisfies  
   all constraints from low energy data and cosmology.
\section*{Acknowledgements}
H.A would like to thank the organisers of the Moriond Conference for the
 invitation to give this talk and for financial assistance. H.A would like
  to thank M.~Plumacher for helpful discussions. 

 

\section*{References} 
\begin{thebibliography}{99} 
\bibitem{WMAP} D.~N. Spergel et~al. \newblock {Astrophys.J.Suppl.}, 148: 175, 2003.
\bibitem{sakharov} A.D. Sakharov. \newblock {\em JETP Lett.}, 5:24, 1967.
\bibitem{smcp} M.B.~Gavela {\it et al}. Nucl.Phys.B430:382-426,1994~; 
Nucl.Phys.B430:345-381,1994.
\bibitem{superk}  {Super-Kamiokande Collaboration (Y. Fukuda {\it et al.})}, 
 {Phys. Rev. Lett. 81}, 1998, 1562. 
 \bibitem{sno} {SNO Collaboration (Q.R. Ahmed {\it et al.})}, { Phys. Rev. Lett. 89}, 2002, 011301. 
 
\bibitem{FY} M.~Fukugita and T.~Yanagida. \newblock {\em Phys. Lett.}, B174:45, 1986.
\bibitem{seesaw} {P.~Ramond}. {hep-ph/9809459}; {T.~Yanagida.} KEK, 1979; 
{M.~Gell-Mann, P.~Ramond, and R.~Slansky}. {Supergravity}, North-Holland Amsterdam, 
1979; {R.N.~Mohapatra, G.~Senjanovic.}{Phys. Rev. Lett.}, 
44:912, 1980; {\sc E.~Witten,}{\em Phys. Lett. B91}, 1980, p.~81.
\bibitem{roulet} L.~Covi, E.~Roulet, and F.~Vissani.  \newblock {\em Phys. Lett.}, 
B384:169--174, 1996.
\bibitem{kuzmin} V.A. Kuzmin, V.A. Rubakov, and M.E. Shaposhnikov. 
\newblock{\em Phys. Lett.}, B155:36, 1985.
\bibitem{bdp1} W.~Buchmuller, P.~Di~Bari, and M.~Plumacher. {Nucl. Phys.},
 B643:367--390, 2002.
\bibitem{di} S.~Davidson and A.~Ibarra.  {Phys. Lett.}, B535:25-32, 2002.
\bibitem{hambye} T.~Hambye.  {\em Nucl. Phys.}, B633:171-192, 2002.
\bibitem{Senami} M.~Senami and K.~Yamamoto. {hep-ph/0305202}, 2003. 

\bibitem{Senami2} M.~Senami and K.~Yamamoto. {Phys.Rev. D69}, 035004, 2004.

\bibitem{Boubekeur} L.~Boubekeur. { hep-ph/0208003}, 2002.

\bibitem{asm} A.~Abada, and M.~Losada. {\em Nucl. Phys.}, B673:319-330, 2003.
\bibitem{pdg} Particle~Data Group. \newblock {\em Euro. Phys. J.}, C15:1, 2002.

\bibitem{kolb} E.W.~Kolb, S.~Wolfram, Nucl.Phys. B172:224, 1980, Erratum-ibid. 
B195:542,1982. 

\bibitem{plumacher} M. Plumacher, Nucl.Phys.B530:207-246,1998. 

\bibitem{branco} G.C. Branco, {\it et al.},  Phys. Rev. D67: 073025, 2003.

 \bibitem{yanagida} J.~Sato and T.~Yanagida.  \newblock {Phys. Lett.}, B430:127--131,
  1998.
\bibitem{Ramond2} N.~Irges, S.~Lavignac, and P.~Ramond.
\newblock {Phys. Rev.}, D58:035003, 1998.
\end{thebibliography}
\end{document}